\documentclass[journal]{IEEEtran}

\usepackage{graphicx}
\usepackage{amsmath}
\usepackage{amssymb}
\usepackage{amsbsy}

\begin{document}
%
\title{Detection of Glottal Closure Instants from Speech Signals: a Quantitative Review}

\author{Thomas Drugman, Mark Thomas, Jon Gudnason, Patrick Naylor, Thierry Dutoit}%



\maketitle

\begin{abstract}
The pseudo-periodicity of voiced speech can be exploited in several speech processing applications. This requires however that the precise locations of the Glottal Closure Instants (GCIs) are available. The focus of this paper is the evaluation of automatic methods for the detection of GCIs directly from the speech waveform.  Five state-of-the-art GCI detection algorithms are compared using six different databases with contemporaneous electroglottographic recordings as ground truth, and containing many hours of speech by multiple speakers. The five techniques compared are the Hilbert Envelope-based detection (HE), the Zero Frequency Resonator-based method (ZFR), the Dynamic Programming Phase Slope Algorithm (DYPSA), the Speech Event Detection using the Residual Excitation And a Mean-based Signal (SEDREAMS) and the Yet Another GCI Algorithm (YAGA). The efficacy of these methods is first evaluated on clean speech, both in terms of reliabililty and accuracy. Their robustness to additive noise and to reverberation is also assessed. A further contribution of the paper is the evaluation of their performance on a concrete application of speech processing: the causal-anticausal decomposition of speech. It is shown that for clean speech, SEDREAMS and YAGA are the best performing techniques, both in terms of identification rate and accuracy. ZFR and SEDREAMS also show a superior robustness to additive noise and reverberation.
\end{abstract}


%
\IEEEpeerreviewmaketitle


\let\thefootnote\relax\footnotetext{T. Drugman and T. Dutoit are with the TCTS Lab, University of Mons, Belgium. M. Thomas and P. Naylor are with the Imperial College of London, UK. J. Gudnason is with the University of Iceland.}

\section{Introduction}\label{sec:Intro}
\IEEEPARstart{G}{lottal}-synchronous speech processing is a field of speech science in which the pseudoperiodicity of voiced speech is exploited. Research into the tracking of pitch contours has proven useful in the field of phonetics~\cite{Catford1977} and speech quality assessment~\cite{ITU_T_P862}; however more recent efforts in the detection of Glottal Closure Instants (GCIs) enable the estimation of both pitch contours and, additionally, the boundaries of individual cycles of speech. Such information has been put to practical use in applications including prosodic speech modification~\cite{Moulines1990}, speech dereverberation~\cite{Gaubitch2007}, glottal flow estimation \cite{Wong1979}, speech synthesis \cite{HNM}, \cite{DSM}, data-driven voice source modelling~\cite{Thomas2009} and causal-anticausal deconvolution of speech signals~\cite{MixedPhase}.

Increased interest in glottal-synchronous speech processing has brought about a corresponding demand for automatic and reliable detection of GCIs from both clean speech and speech that has been corrupted by acoustic noise sources and/or reverberation. Early approaches that search for maxima in the autocorrelation function of the speech signal~\cite{Strube1974a} were found to be unreliable due to formant frequencies causing multiple maxima. More recent methods search for discontinuities in the linear production model of speech~\cite{Rabiner1988} by deconvolving the excitation signal and vocal tract filter with linear predictive coding (LPC)~\cite{Makhoul1975}. Preliminary efforts are documented in~\cite{Wong1979}; more recent algorithms use known features of speech to achieve more reliable detection~\cite{Plumpe1999, Naylor2007a,Thomas2010b}. Deconvolution of the vocal tract and excitation signal by homomorphic processing~\cite{Chytil2006} has also been used for GCI detection although its efficacy compared with LPC has not been fully researched. Various studies have shown that, while linear model-based approaches can give accurate results on clean speech, reverberation can be particularly detrimental to performance~\cite{Gaubitch2007, Thomas2007a}.

Methods that use smoothing or measures of energy in speech signal are also common. These include the Hilbert Envelope~\cite{Ananthapadmanabha1979}, Frobenius Norm~\cite{Ma1994}, Zero-Frequency Resonator (ZFR)~\cite{Murty2008} and SEDREAMS~\cite{SEDREAMS}. Smoothing of the speech signal is advantageous because the vocal tract resonances, additive noise and reverberation are attenuated while the periodicity of the speech signal is preserved. A disadvantage lies in the ambiguity of the precise time instant of the GCI; for this reason LP residual can be used in addition to smoothed speech to obtain more accurate estimates~\cite{Naylor2007a, SEDREAMS}. Smoothing  on multiple dyadic scales is exploited by wavelet decomposition of the speech signal with the Multiscale Product~\cite{Bouzid2004} and Lines of Maximum Amplitudes (LOMA)~\cite{Tuan1999a} to achieve both accuracy and robustness. The YAGA algorithm~\cite{Thomas2010b} employs both multiscale processing and the linear speech model.

The aim of this paper is to provide a review and objective evaluation of five contemporary methods for GCI detection, namely Hilbert Envelope-based method~\cite{Ananthapadmanabha1979}, DYPSA~\cite{Naylor2007a}, ZFR~\cite{Murty2008}, SEDREAMS~\cite{SEDREAMS} and YAGA~\cite{Thomas2010b} algorithms. In their corresponding references, all these techniques reported interesting results and were shown to outperform other state-of-the-art methods. Besides, as described in Section \ref{sec:Methods}, they rely on different approaches: some are based on the speech signal while others focus on the residual signal or an estimate of the glottal source; some use dynamic programming while others exploit a smoothing process. As a consequence, these techniques may have different properties in terms of reliability, accuracy and robustness. They are here evaluated against reference GCIs provided by an Electroglottograph (EGG) signal on six databases, of combined duration 232 minutes, containing contemporaneous recordings of EGG and speech. Performance is also evaluated in the presence of additive noise and reverberation. A novel contribution of this paper is the application of the algorithms to causal-anticausal deconvolution~\cite{MixedPhase}, which provides additional insight into their performance in a real-world problem.

The remainder of this paper is organised as follows. In Section~\ref{sec:Methods} the algorithms under test are described. In Section~\ref{sec:Assessment} the evaluation techniques are described. Sections~\ref{sec:ExpClean} and~\ref{sec:Robustness} discuss the performance results on clean and noisy/reverberant speech respectively. Section~\ref{sec:Complexity} compares the methods in terms of computational complexity. Conclusions are given in Section~\ref{sec:conclu}.

\section{Methods Compared in this Work}\label{sec:Methods}

This Section presents five of the main representative state-of-the-art methods for automatically detecting GCIs from speech waveforms. These techniques are detailed here below and their reliability, accuracy and robustness will be compared in Sections \ref{sec:ExpClean} and \ref{sec:Robustness}. It is worth noting at this point that all methods assume a positive polarity of the speech signal. Polarity should then be verified and corrected if required, using an algorithm such as \cite{Polarity}.

\subsection{Hilbert Envelope-based method}\label{ssec:HE}
Several approaches relying on the Hilbert Envelope (HE) have been proposed in the literature \cite{Yegnanarayana1979,Cheng1989,Rao2007}. In this article, a method based on the HE of the Linear Prediction (LP) residual signal (i.e the signal whitened by inverse filtering after removing an auto-regressive modeling of the spectral envelope) is considered. 

\begin{figure}[!ht]
  \centering
  \includegraphics[width=0.5\textwidth]{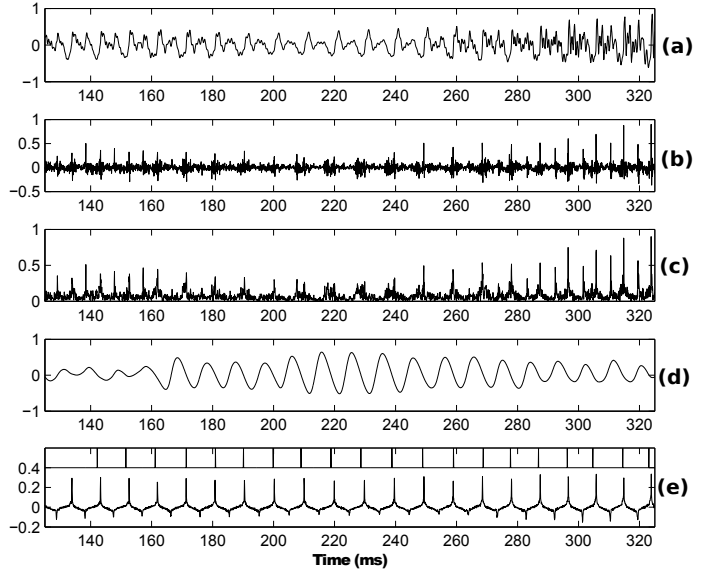}
  \caption{Illustration of GCI detection using the Hilbert Envelope-based method on a segment of voiced speech. \emph{(a) :} the speech signal, \emph{(b) :} the LP residual signal, \emph{(c) :} the Hilbert Envelope (HE) of the LP residue, \emph{(d) :} the Center of Gravity-based signal computed from the HE, \emph{(e) :} the synchronized differenced EGG with the GCI positions located by the HE-based method.}
  \label{fig:HE_illus}
\end{figure}

Figure \ref{fig:HE_illus} illustrates the principle of this method for a short segment of voiced speech (Fig.\ref{fig:HE_illus}(a)). The corresponding synchronized derivative of the ElectroGlottoGraph (dEGG) is displayed in Fig.\ref{fig:HE_illus}(e), as it is informative about the actual positions of both GCIs (instants where the dEGG has a large positive value) and GOIs (instants of weaker negative peaks between two successive GCIs). The LP residual signal (shown in Fig.\ref{fig:HE_illus}(b)) contains clear peaks around the GCI locations. Indeed the impulse-like nature of the excitation at GCIs is reflected by discontinuities in this signal. It is also observed that for some glottal cycles (particularly before 170 ms or beyond 280 ms) the LP residual also presents clear discontinuities around GOIs. The resulting HE of the LP residual, containing large positive peaks when the excitation presents discontinuities, and its Center of Gravity (CoG)-based signal are respectively exhibited in Figures \ref{fig:HE_illus}(c) and \ref{fig:HE_illus}(d). Denoting $H_e(n)$ the Hilbert envelope of the residue at sample index $n$, the CoG-based signal is defined as:

\begin{equation}\label{eq:CoG}
CoG(n)=\frac{\sum_{m=-N}^N{m \cdot w(m)H_e(n+m)}}{\sum_{m=-N}^N{w(m)H_e(n+m)}}	
\end{equation}

where $w(m)$ is a windowing function of length $2N+1$. In this work a Blackman window whose length is 1.1 times the mean pitch period of the considered speaker was used. We empirically reported in our experiments that using this window length led to a good compromise between misses and false alarms (i.e to the best reliability performance). Once the CoG-based signal is computed, GCI locations correspond to the instants of negative zero-crossing. The resulting GCI positions obtained for the speech segment are indicated in the top of Fig.\ref{fig:HE_illus}(e). It is clearly noticed that the possible ambiguity with the discontinuities around GOIs is removed by using the CoG-based signal. 

\subsection{The DYPSA algorithm}\label{ssec:DYPSA}
The Dynamic Programming Phase Slope Algorithm (DYPSA)~\cite{Naylor2007a} estimates GCIs by the identification of peaks in the linear prediction residual of speech in a similar way to the HE method. It consists of two main components: estimation of GCI candidates with the group delay function of the LP residual and $N$-best dynamic programming. These components are defined as follows.

\subsubsection{Group Delay Function}\label{sssec:GD} 
The group delay function is the average slope of the unwrapped phase spectrum of the short time Fourier transform of the LP residual~\cite{Yegnanarayana1995}~\cite{Brookes2006}. It can be shown to accurately identify impulsive features in a function provided their minimum separation is known. GCI candidates are selected based on the negative-going zero crossings of the group delay function. Consider an LP residual signal, $e(n)$, and an $R$-sample windowed segment $x_n(r)$ beginning at sample $n$
\begin{equation}
    x_n(r)=w(r)e(n+r)~\text{for}~r=0,\dots,R-1
\end{equation}
where $w(r)$ is a windowing function. The group delay of $x_n(r)$ is given by~\cite{Yegnanarayana1995}
\begin{equation}\label{eq:grdel}
    \tau_n(k)=\frac{-\text{d}\arg(X_n)}{\text{d}\omega}=\Re\left(\frac{\tilde{X}_n(k)}{X_n(k)}\right)
\end{equation}
where $X_n(k)$ is the Fourier transform of $x_n(r)$ and  $\tilde{X}_n(k)$ is the Fourier transform of $rx_n(r)$.
If $x_n(r)=\delta(r-r_0)$, where $\delta(r)$ is a unit impulse function, it follows from (\ref{eq:grdel}) that $\tau_n(k)\equiv r_0\forall k$. In the presence of noise, $\tau_n(k)$ becomes noisy, therefore an averaging procedure is performed over $k$. Different approaches are reviewed in~\cite{Brookes2006}. The \textit{Energy-Weighted Group Delay} is defined as
\begin{equation}
    d(n)=\frac{\sum^{R-1}_{k=0}|X_n(k)|^2\tau_n(k)}{\sum^{R-1}_{k=0}|X_n(k)|^2}-\frac{R-1}{2}.
\end{equation}
Manipulation yields the simplified expression
\begin{equation}\label{eq:ew_grdel}
    d(n)=\frac{\sum^{R-1}_{r=0}rx^2_n(r)}{\sum^{R-1}_{r=0}x^2_n(r)}-\frac{R-1}{2}
\end{equation}
which is an efficient time-domain formulation and can be viewed as a centre of gravity of $x_n(r)$, bounded in the range $[-(R-1)/2, (R-1)/2]$. The location of the negative-going zero crossings of $d(n)$ give an accurate estimation of the location of a peak in a function.

It can be shown that the signal $d(n)$ does not always produce a negative-going zero crossing when an impulsive feature occurs in $e(n)$. In such cases, it has been observed that $d(n)$ consistently exhibits local minima followed by local maxima in the vicinity of the impulsive feature~\cite{Naylor2007a}. A \emph{phase-slope projection} technique is therefore introduced to estimate the time of the impulsive feature by finding the midpoint between local maxima and minima where no zero crossing is produced, then projecting a line onto the time axis with negative unit slope.

\subsubsection{Dynamic Programming}
Erroneous GCI candidates are removed using known characteristics of voiced speech by minimising a cost function so as to select a subset of the GCI candidates which most likely correspond to true GCIs. The subset of candidates is selected according by minimising the following cost function
\begin{equation}
\min_\Omega\sum_{r=1}^{|\Omega|}\boldsymbol{\lambda}^T\textbf{c}_\Omega(r),
\end{equation}
where $\Omega$ is a subset with GCI candidates of size $|\Omega|$ selected to produce minimum cost,  $\boldsymbol{\lambda}=[\lambda_{A}~\lambda_{P}~\lambda_{J}~\lambda_{F}~\lambda_{S}]^T=[0.8~0.5~0.4~0.3~0.1]^T$ is a vector of weighting factors, the choice of which is described in \cite{Naylor2007a}, and $\textbf{c}(r)=[c_A(r)~c_P(r)~c_J(r)~c_F(r)~c_S(r)]^T$ is a vector of cost elements evaluated at the $r$th element of $\Omega$. The cost vector elements are:
\begin{itemize}
		\item \emph{Speech waveform similarity}, $c_A(r)$, between neighbouring candidates, where candidates not correlated with the previous candidate are penalised.
		
		\item \emph{Pitch deviation}, $c_P(r)$, between the current and the previous two candidates, where candidates with large deviation are penalised.
		
		\item \emph{Projected candidate cost}, $c_J(r)$, for the candidates from the phase-slope projection, which often arise from erroneous peaks.
		
		\item \emph{Normalised energy}, $c_F(r)$, which penalises candidates that do not correspond to high energy in the speech signal.
		
		\item \emph{Ideal phase-slope function deviation}, $c_S(r)$, where candidates arising from zero-crossings with gradients close to unity are favoured.
	\end{itemize}

\subsection{The Zero Frequency Resonator-based technique}\label{ssec:ZFR}

The Zero Frequency Resonator-based (ZFR) technique relies on the observation that the impulsive nature of the excitation at GCIs is reflected across all frequencies \cite{Murty2008}.  The GCI positions can be detected by confining the analysis around a single frequency.  More precisely, the method focuses the analysis on the output of zero frequency resonators to guarantee that the influence of vocal-tract resonances is minimal and, consequently, that the output of the zero frequency resonators is mainly controlled by the excitation pulses. The zero frequency-filtered signal (denoted $y(n)$ here below) is obtained from the speech waveform $s(n)$ by the following operations \cite{Murty2008}:

\begin{enumerate}
\item Remove from the speech signal the dc or low-frequency bias during recording:

\begin{equation}
x(n)=s(n)-s(n-1)
\end{equation}

\item Pass this signal two times through an ideal zero-frequency resonator:

\begin{equation}
y_1(n)=x(n)+2\cdot y_1(n-1)+y_1(n-2)
\end{equation}

\begin{equation}
y_2(n)=y_1(n)+2\cdot y_2(n-1)+y_2(n-2)
\end{equation}

The two passages are necessary for minimizing the influence of the vocal tract resonances in $y_2(n)$.

\item As the resulting signal $y_2(n)$ is exponentially increasing or decreasing after this filtering, its trend is removed by a mean-substraction operation:

\begin{equation}\label{eq:MeanRemoval}
y(n)=y_2(n)-\frac{1}{2N+1}\sum_{m=-N}^N{y_2(n+m)}	
\end{equation}

where the window length $2N+1$ was reported in \cite{Murty2008} to be not very critical, as long as it is in the
range of about 1 to 2 times the average pitch period $\bar{T}_{0,mean}$ of the considered speaker. Accordingly, we used in this study a window whose length is 1.5$\cdot$$\bar{T}_{0,mean}$. Note also that this operation of mean removal has to be repeated three times in order to avoid any residual drift of $y(n)$.
\end{enumerate}

\begin{figure}[!ht]
  \centering
  \includegraphics[width=0.5\textwidth]{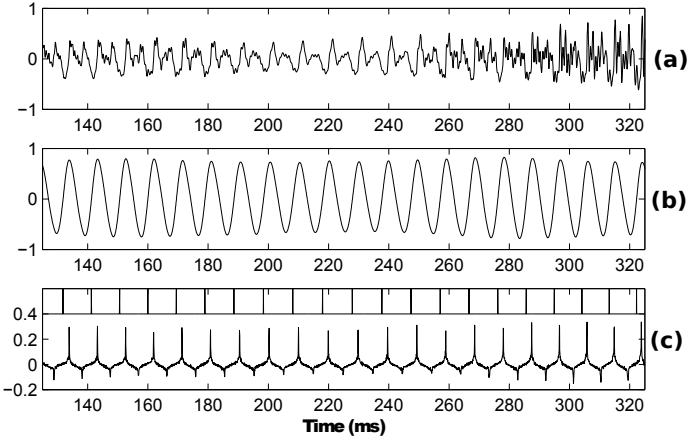}
  \caption{Illustration of GCI detection using the Zero Frequency Resonator-based method on a segment of voiced speech. \emph{(a) :} the speech signal, \emph{(b) :} the zero frequency-filtered signal, \emph{(c) :} the synchronized dEGG with the GCI positions located by the ZFR-based method.}
  \label{fig:ZFR_illus}
\end{figure}

An illustration of the resulting zero frequency-filtered signal is displayed in Fig. \ref{fig:ZFR_illus}(b) for our example. This signal is observed to possess two advantageous properties: 1) it oscillates at the local pitch period, 2) the positive zero-crossings of this signal correspond to the GCI positions. This is confirmed in Fig. \ref{fig:ZFR_illus}(c), where a good agreement is noticed between the GCI locations identified by the ZFR technique and the actual discontinuities in the synchronized dEGG.

\subsection{The SEDREAMS algorithm}\label{ssec:SEDREAMS}
The Speech Event Detection using the Residual Excitation And a Mean-based Signal (SEDREAMS) algorithm was recently proposed in \cite{SEDREAMS} as a reliable and accurate method for locating both GCIs and GOIs from the speech waveform. Since the present study only focuses on GCIs, the determination of GOI locations by the SEDREAMS algorithm is omitted. The two steps involved in this method are: \emph{i)} the determination of short intervals where GCIs are expected to occur and \emph{ii)} the refinement of the GCI locations within these intervals. These two steps are described in the following subsections.

\subsubsection{Determining intervals of presence using a mean-based signal}\label{sssec:Mean-based}

As highlighted by the ZFR technique \cite{Murty2008}, a discontinuity in the excitation is reflected over the whole spectral band, including the zero frequency. Inspired by this observation, the analysis is focused on a mean-based signal. Denoting the speech waveform as $s(n)$, the mean-based signal $y(n)$ is defined as:

\begin{equation}\label{eq:Mean}
y(n)=\frac{1}{2N+1}\sum_{m=-N}^N{w(m)s(n+m)}	
\end{equation}

where $w(m)$ is a windowing function of length $2N+1$. While the choice of the window shape is not critical (a typical Blackman window is used in this study), it has been shown \cite{SEDREAMS} that its length, which influences the time response of this filtering operation, may affect the reliability of the method.

\begin{figure}[!ht]
  \centering
  \includegraphics[width=0.45\textwidth]{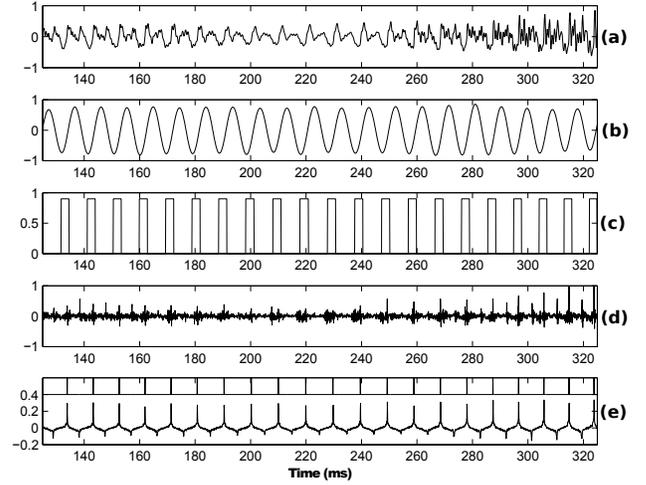}
  \caption{Illustration of GCI detection using the SEDREAMS algorithm on a segment of voiced speech. \emph{(a) :} the speech signal, \emph{(b) :} the mean-based signal, \emph{(c) :} intervals of presence derived from the mean-based signal, \emph{(d) :} the LP residual signal, \emph{(e) :} the synchronized dEGG with the GCI positions located by the SEDREAMS algorithm.}
  \label{fig:SEDREAMS_illus}
\end{figure}

A segment of voiced speech and its corresponding mean-based signal using an appropriate window length are illustrated in Figs. \ref{fig:SEDREAMS_illus}(a) and \ref{fig:SEDREAMS_illus}(b). Interestingly it is observed that the mean-based signal oscillates at the local pitch period. If the window is too short, it causes the appearance of spurious extrema in the mean-based signal, giving rise to false alarms. On the other hand, too large a window smooths it, leading to some possible misses. It has been observed in \cite{SEDREAMS} that maximal reliability is obtained when the window length is between 1.5 and 2 times the average pitch period $\bar{T}_{0,mean}$ of the considered speaker. Accordingly, throughout the rest of this article a window whose length is 1.75$\cdot$$\bar{T}_{0,mean}$ is used for computing the mean-based signal of the SEDREAMS algorithm. 

\begin{figure}[!ht]
  \centering
  \includegraphics[width=0.45\textwidth]{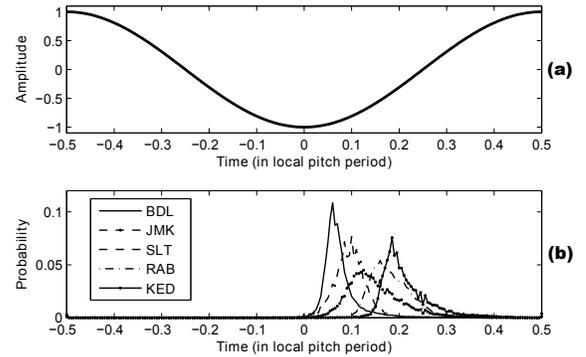}
  \caption{Distributions, for five speakers, of the actual GCI positions (plot \emph{(b)}) within a normalized cycle of the mean-based signal (plot \emph{(a)}).}
  \label{fig:RelativePosition}
\end{figure}

However the mean-based signal is not sufficient in itself for accurately locating GCIs. Indeed, consider Fig. \ref{fig:RelativePosition} where, for five different speakers, the distributions of the actual GCI positions (extracted from synchronized EGG recordings) are displayed within a normalized cycle of the mean-based signal. It turns out that GCIs may occur at a non-constant relative position within the cycle. However, once minima and maxima of the mean-based signal are located, it is straightforward to derive short intervals of presence where GCIs are expected to occur. More precisely, as observed in Fig. \ref{fig:RelativePosition}, these intervals are defined as the timespan starting at the minimum of the mean-based signal, and whose length is 0.35 times the local pitch period (i.e the period between two consecutive minima). Such intervals are illustrated in Fig.\ref{fig:SEDREAMS_illus}(c) for our example.

\subsubsection{Refining GCI locations using the residual excitation}\label{sssec:Refinement}
Intervals of presence obtained in the previous step give fuzzy short regions where a GCI should happen. The goal of the next step is to refine, for each of these intervals, the precise location of the GCI occuring inside it. The LP residual is therefore inspected, assuming that the largest discontinuity of this signal within a given interval corresponds to the GCI location.

Figs. \ref{fig:SEDREAMS_illus}(d) and \ref{fig:SEDREAMS_illus}(e) show the LP residual and the time-aligned dEGG for our example. It is clearly noted that combining the intervals extracted from the mean-based signal with a peak picking method on the LP residue allows the accurate and unambiguous detection of GCIs (as indicated in Fig.\ref{fig:SEDREAMS_illus}(e)).

It is worth noting that the advantage of using the mean-based signal is two-fold. First of all, since it oscillates at the local pitch period, this signal guarantees good performance in terms of reliability (i.e the risk of misses or false alarms is limited). Secondly, the intervals of presence that are derived from this signal imply that the GCI timing error is bounded by the depth of these intervals (i.e 0.35 times the local pitch period).

\subsection{The YAGA algorithm}\label{ssec:YAGA}

\begin{figure}[!ht]
  \centering
  \includegraphics[width=0.45\textwidth]{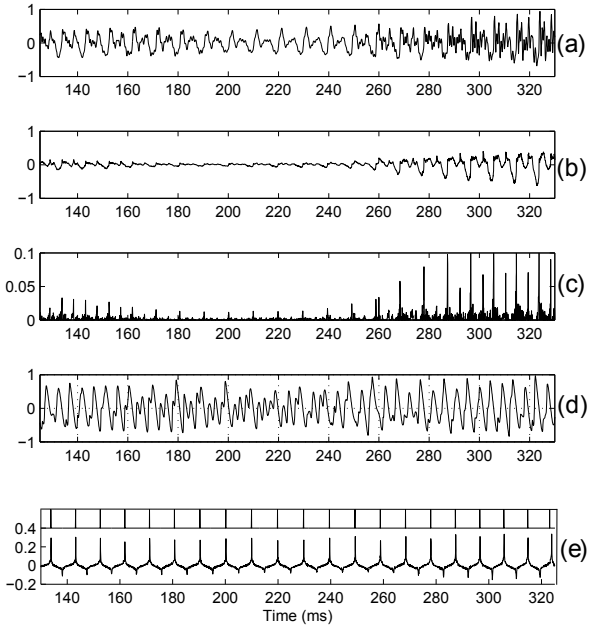}
  \caption{Illustration of GCI detection using the YAGA algorithm on a segment of voiced speech. \emph{(a) :} the speech signal, \emph{(b) :} the corresponding voice source signal, \emph{(c) :} the multiscale product of the voice source, \emph{(d) :} the group-delay function, \emph{(e) :} the synchronized dEGG with the GCI positions located by the YAGA algorithm.}
  \label{fig:YAGA}
\end{figure}

The Yet Another GCI Algorithm (YAGA)~\cite{Thomas2010b}, like DYPSA, is an LP-based approach that employs $N$-best dynamic programming to find the best path through a set of candidate GCIs. The algorithms differ in the way in which the candidate set is estimated. Candidates are derived in DYPSA using a linear prediction residual, calculated by inverse-filtering a preemphasised speech signal with the LP coefficients. GCIs are manifest as impulsive features that may be detected with the group delay function. In YAGA, candidates are derived from an estimate of the voice source signal $u'(n)$ by using the same LP coefficients to inverse-filter the non-preemphasized speech signal. This differs crucially in that it exhibits discontinuities at both GCIs and GOIs, although GOIs are not considered in this paper. The speech signal $s(n)$ and voice source signal $u'(n)$ are shown for a short speech sample in Fig.~\ref{fig:YAGA} (a) and (b) respectively.

The impulsive nature of the LPC residual is well-suited to detection with the group delay method as discussed in Section~\ref{ssec:DYPSA}. In order for the group delay method to be applied to voice source signal, a discontinuity detector that yields an impulse-like signal is required. Such a detector might be achieved by a 1st-order differentiator, however it is known that GCIs and GOIs are not instantaneous discontinuities but are instead spread over time~\cite{Bouzid2004}. The Stationary Wavelet Transform (SWT) is a multiscale analysis tool for the detection of discontinuities in a signal by considering the product of the signal at different scales~\cite{Mallat1992}. It was first used in the context of GCI detection in~\cite{Bouzid2004} by application to the speech signal. YAGA employs a similar approach on the voice source signal, which is expected to yield better results as it is free from unwanted vocal tract resonances. The SWT of signal $u'(n)$, $1\leq n\leq N$ at scale $j$ is
\begin{align}
	d^s_j(n)&=W_{2^j}u'(n), \nonumber \\ 
		&=\sum_{k}g_j(k)a_{j-1}^s(n-k),
\end{align}
where the maximum scale $J$ is bounded by $\log_2N$ and $j=1,2,\dots,J-1$. The approximation coefficients are given by
\begin{equation}
	a^s_j(n)=\sum_{k}h_j(k)a_{j-1}^s(n-k),
\end{equation}
where $a^s_0(n)=u'(n)$ and $g_j(k)$, $h_j(k)$ are detail and approximation filters respectively that are upsampled by two on each iteration to effect a change of scale~\cite{Mallat1992}. Filters are derived from a biorthogonal spline wavelet with one vanishing moment~\cite{Mallat1992}.
The multiscale product, $p(n)$, is formed by
\begin{equation}
	p(n)=\prod^{j_1}_{j=1}d_j(n) = \prod^{j_1}_{j=1}W_{2^j}u'(n),
\end{equation}
where it is assumed that the lowest scale to include is always 1. The de-noising effect of the approximation filters each scale in conjunction with the multiscale product means that $p(n)$ is near-zero except at discontinuities across the first $j_1$ scales of $u'(n)$ where it becomes impulse-like. The value of $j_1$ is bounded by $J$, but in practice $j_1=3$ gives good localization of discontinuities in acoustic signals~\cite{Sadler1999}. 

The multiscale product of the voice source signal in Fig.~\ref{fig:YAGA} (b) is shown in plot (c). Impulse-like features can be seen in the vicinity of discontinuities of $u'(n)$; such features are then detected by the negative-going zero-crossings of the group delay function in plot (d) that  form the candidate set of GCIs. In order to distinguish between GCIs, GOIs and false candidates, an $N$-best dynamic programming algorithm is applied. The cost function employed is similar to that of DYPSA with an improved waveform similarity measure and an additional element to reliably differentiate between GCIs and GOIs.



\section{Assessment of GCI Extraction Techniques}\label{sec:Assessment}

\subsection{Speech Material}\label{ssec:Material}
The evaluation of the GCI detection methods relies on ground-truth obtained from EGG recordings. The methods are compared on six large corpora containing contemporaneous EGG recordings whose description is summarized in Table~ \ref{tab:TabDba}. The first three corpora come from the CMU ARCTIC databases \cite{Festvox}. They were collected at the Language Technologies Institute at Carnegie Mellon University with the goal of developing unit selection speech synthesizers. Each phonetically balanced dataset contains 1150 sentences uttered by a single speaker: BDL (US male), JMK (US male) and SLT (US female).  The fourth corpus consists of a set of nonsense words containing all phone-phone transitions for English, uttered by the UK male speaker RAB.  The fifth corpus is the KED Timit database and contains 453 utterances spoken by a US male speaker. These five first databases are freely available on the Festvox webpage \cite{Festvox}.  The sixth corpus is the APLAWD dataset \cite{Lindsey1987} which contains ten repetitions of five phonetically balanced English sentences spoken by each of five male and five female talkers. For each of these six corpora, the speech and EGG signals sampled at 16 kHz are considered. The APLAWD database contains a square wave calibration signal for correcting low-frequency phase distortion, introduced in the recording chain, with an allpass equalization filter~\cite{Hunt1978}. While this is particularly important in the field of voice source estimation and modelling~\cite{Funaki1999}, we have found GCI detection to be relatively insensitive to such phase distortion. An intuitive explanation is that the glottal excitation at the GCI excites many high-frequency bins such that low-frequency distortion does not have a significant effect upon the timing of the estimated GCI.

\begin{table}[!ht]
\centering
\begin{tabular}{| c | c | c | c |}
\hline
\textbf{Dataset} & \textbf{Speaker(s)} & \textbf{Approximative duration}\\
\hline
BDL & 1 male & 54 min.\\
\hline
JMK & 1 male & 55 min.\\
\hline
SLT & 1 female & 54 min.\\
\hline
RAB & 1 male & 29 min.\\
\hline
KED & 1 male & 20 min.\\
\hline
APLAWD & 5 males - 5 females & 20 min.\\
\hline
\hline
Total & 9 males - 6 females & 232 min.\\
\hline
\end{tabular}
\caption{Description of the databases.}
\label{tab:TabDba}
\end{table}

\subsection{Objective Evaluation}\label{ssec:Evaluation}
The most common way to assess the performance of GCI detection techniques is to compare the estimates with the reference locations extracted from EGG signals (Section \ref{sssec:compEGG}). Besides it is also proposed to evaluate their efficiency on a specific application of speech processing: the causal-anticausal deconvolution (Section \ref{sssec:MixedPhase}).

\subsubsection{Comparison with Electroglottographic Signals}\label{sssec:compEGG}
Electroglottography (EGG), also known as electrolaryngography, is a non-intrusive technique for measuring the time-varying impedance between the vocal folds. The EGG signal is obtained by passing a weak electrical current between a pair of electrodes placed in contact with the skin on both sides of the larynx. This measure is proportionate to the contact area of the vocal folds. As clearly seen in the explanatory figures of Section \ref{sec:Methods}, true positions of GCIs can then be easily detected by locating the greatest positive peaks in the differenced EGG signal. Note that, for the automatic assessment, EGG signals need to be time-aligned with speech signals by compensating the delay between the EGG and the microphone. This was done in this work by a manual verification for each database (inside which the delay is assumed to remain constant).

\begin{figure}[!ht]
  \centering
  \includegraphics[width=0.45\textwidth]{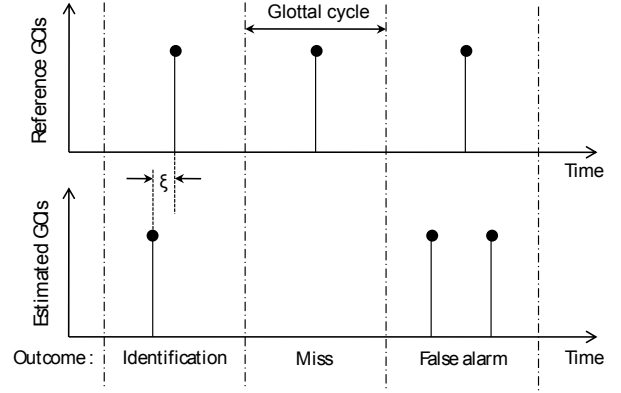}
  \caption{Characterization of GCI estimates showing three glottal cycles with examples of each possible outcome from GCI estimation \cite{Naylor2007a}. Identification accuracy is characterized by $\xi$.}
  \label{fig:ErrorMeasures}
\end{figure}

Performance of a GCI detection method can be evaluated by comparing the locations that are estimated with the synchronized reference positions derived from the EGG recording. For this, we here make use of the performance measure defined in \cite{Naylor2007a}, presented with the help of Fig. \ref{fig:ErrorMeasures}. The first three measures describe how \emph{reliable} the algorithm is in identifying GCIs:

\begin{itemize}
\item the Identification Rate (IDR): the proportion of glottal cycles for which exactly one GCI is detected,
\item the Miss Rate (MR): the proportion of glottal cycles for which no GCI is detected,
\item and the False Alarm Rate (FAR): the proportion of glottal cycles for which more than one GCI is detected.
\end{itemize}

For each correct GCI detection (i.e respecting the IDR criterion), a timing error $\xi$ is made with reference to the EGG-derived GCI position. When analyzing a given dataset with a particular method of GCI detection, $\xi$ has a probability density comparable to the histograms of Fig. \ref{fig:ErrorHisto} (which will be detailed later in this paper). Such a distribution can be characterized by the following measures for quantifying the \emph{accuracy} of the method \cite{Naylor2007a}:

\begin{itemize}
\item the Identification Accuracy (IDA): the standard deviation of the distribution,
\item the Accuracy to $\pm$ 0.25 ms: the proportion of detections for which the timing error is smaller than this bound.
\end{itemize}

\subsubsection{A Speech Processing Application: the Causal-Anticausal Deconvolution}\label{sssec:MixedPhase}
The causal-anticausal decomposition (also known as mixed-phase decomposition) is a non-parametric technique of source-tract deconvolution known to be highly sensitive to GCI location errors \cite{MixedPhase}. It can therefore be employed as a framework for assessing our methods of GCI extraction on a speech processing application. The principle of this decomposition relies on the mixed-phase model of speech \cite{Doval-CALM}, \cite{MixedPhase}. According to this model, voiced speech is composed of both minimum-phase (i.e causal) and maximum-phase (i.e anticausal) components. While the vocal tract response and the glottal \emph{return phase} can be considered as minimum-phase signals, it has been shown \cite{Doval-CALM} that the glottal \emph{open phase} is a maximum-phase signal. The key idea of the causal-anticausal (or mixed-phase) decomposition is then to separate both minimum and maximum-phase components of speech, where the latter is only due to the glottal contribution. By isolating the anticausal component of speech, causal-anticausal separation allows to estimate the glottal open phase.

\begin{figure}[!ht]
  \centering
  \includegraphics[width=0.45\textwidth]{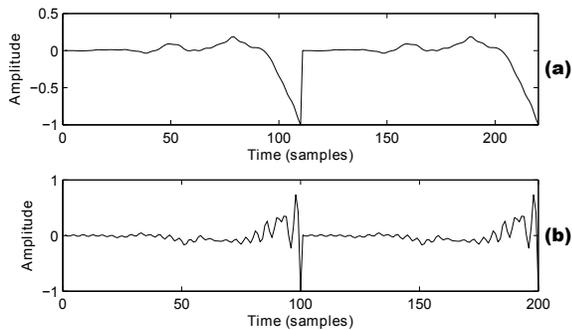}
  \caption{Two cycles of the anticausal component isolated by mixed-phase decomposition \emph(a): when the speech segment exhibits characteristics of the mixed-phase model, \emph(b): when this is not the case.}
  \label{fig:GlottalFlow}
\end{figure}

Two algorithms have been proposed in the literature for achieving the causal-anticausal separation: the Zeros of the Z-Transform (ZZT, \cite{ZZT}) method and the Complex Cepstrum-based Decomposition (CCD, \cite{Drugman-CCD}). It has been shown \cite{Drugman-CCD} that both algorithms are functionally equivalent and lead to a reliable estimation of the glottal flow. However the use of the CCD technique was recommended for its much higher computational speed compared to ZZT. Besides it was also shown in \cite{Drugman-CCD} that windowing is crucial and dramatically conditions the efficiency of the causal-anticausal decomposition. It is indeed essential that the window applied to the segment of voiced speech respects some constraints in order to exhibit correct mixed-phase properties. Among these constraints, the window should be synchronized on a GCI, and have an appropriate shape and length (proportional to the pitch period). If the windowing is such that the speech segment respects the properties of the mixed-phase model, a correct deconvolution is achieved and the anticausal component gives a reliable estimate of the glottal flow (i.e which corroborates the models of the glottal source, such as the LF model \cite{Fant1985}), as illustrated in Fig. \ref{fig:GlottalFlow}(a). On the contrary, if this is not the case (possibly due to the fact that the window is not perfectly synchronized with the GCI), the causal-anticausal decomposition fails, and the resulting anticausal component generally contains an irrelevant high-frequency noise (see Fig.\ref{fig:GlottalFlow}(b)).

\begin{figure}[!ht]
  \centering
  \includegraphics[width=0.45\textwidth]{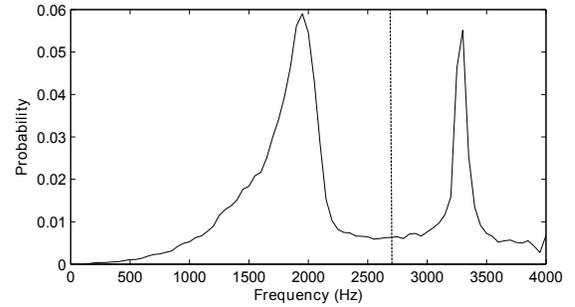}
  \caption{Example of distribution for the spectral center of gravity of the maximum-phase component. Fixing a threshold around 2.7kHz makes a good separation between correctly and incorrectly decomposed frames.}
  \label{fig:COGhisto}
\end{figure}

As a simple (but accurate) criterion for deciding whether a frame has been correctly decomposed or not, the spectral center of gravity of the anticausal component is investigated. For a given dataset, this feature has a distribution as the one displayed in Fig. \ref{fig:COGhisto}. A principal mode around 2 kHz clearly emerges and corresponds to the majority of frames for which a correct decomposition is carried out (as in Fig.\ref{fig:GlottalFlow}(a)). A second mode at higher frequencies is also observed. It is related to the frames where the causal-anticausal decomposition fails, leading to a maximum-phase signal containing an irrelevant high-frequency noise (as in Fig.\ref{fig:GlottalFlow}(b)). It can be noticed from this histogram that fixing a threshold at around 2.7 kHz optimally discriminate frames that are correctly and incorrectly decomposed. 


In conclusion, it is expected that the use of good GCI estimates reduces the proportion of frames that are incorrectly decomposed using the causal-anticausal separation.



\section{Experiments on Clean Speech Data}\label{sec:ExpClean}

Based on the experimental protocol described in Section \ref{sec:Assessment}, the performance of the five methods of GCI detection introduced in Section \ref{sec:Methods} is now compared on the original clean speech utterances.

\subsection{Comparison with Electroglottographic Signals}\label{ssec:compEGG}

Results obtained from the comparison with electroglottographic recordings are presented in Table \ref{tab:TabClean} for the various databases.

\begin{table*}[!ht]
\centering
\begin{tabular}{| c | c || c | c | c || c | c |}
\hline
\textbf{Database} & \textbf{Method} & \textbf{IDR ($\%$)} & \textbf{MR ($\%$)} & \textbf{FAR ($\%$)} & \textbf{IDA (ms)} & \textbf{Accuracy to $\pm 0.25$ms ($\%$)} \\
\hline
\hline
  & HE & 97.04 & 1.93 & 1.03 & 0.58 & 46.24 \\
  & DYPSA & 95.54 & 2.12 & 2.34 & 0.42 & 83.74 \\
BDL  & ZFR & 97.97 & 1.05 & \textbf{0.98} & 0.30 & 80.93 \\
  & SEDREAMS & 98.08 & 0.77 & 1.15 & 0.31 & 89.35 \\
  & YAGA & \textbf{98.43} & \textbf{0.39} & 1.18 & \textbf{0.29} & \textbf{90.31} \\
\hline
  & HE & 93.01 & 3.94 & 3.05 & 0.90 & 38.66 \\
  & DYPSA & 98.26 & 0.88 & 0.86 & 0.46 & 77.26 \\
JMK  & ZFR & 96.17 & 3.43 & \textbf{0.4} & 0.60 & 41.62 \\
  & SEDREAMS & \textbf{99.29} & \textbf{0.25} & 0.46 & 0.42 & 80.78 \\
  & YAGA & 99.13 & 0.27 & 0.60 & \textbf{0.40} & \textbf{81.05} \\
\hline
  & HE & 96.16 & 2.83 & 1.01 & 0.56 & 52.46 \\
  & DYPSA & 97.18 & 1.41 & 1.41 & 0.44 & 72.17 \\
SLT  & ZFR & \textbf{99.26} & 0.15 & \textbf{0.59} & \textbf{0.22} & 83.70 \\
  & SEDREAMS & 99.15 & \textbf{0.12} & 0.73 & 0.30 & 81.35 \\
  & YAGA & 98.90 & 0.20 & 0.90 & 0.28 & \textbf{86.18} \\
\hline
  & HE & 92.08 & 2.55 & 5.37 & 0.78 & 38.67 \\
  & DYPSA & 82.33 & 1.87 & 15.80 & 0.46 & 86.76 \\
RAB  & ZFR & 92.94 & 6.31 & 0.75 & 0.56 & 55.87 \\
  & SEDREAMS & \textbf{98.87} & \textbf{0.63} & \textbf{0.50} & \textbf{0.37} & \textbf{91.26} \\
  & YAGA & 95.70 & 0.47 & 3.83 & 0.49 & 89.77 \\
\hline
  & HE & 94.73 & 1.75 & 3.52 & 0.56 & 65.81 \\
  & DYPSA & 97.24 & 1.56 & 1.20 & 0.34 & 89.46 \\
KED  & ZFR & 87.36 & 7.90 & 4.74 & 0.63 & 46.82 \\
  & SEDREAMS & \textbf{98.65} & 0.67 & \textbf{0.68} & \textbf{0.33} & 94.65 \\
  & YAGA & 98.21 & \textbf{0.63} & 1.16 & 0.34 & \textbf{95.14} \\
\hline
  & HE & 91.74 & 5.64 & 2.62 & 0.73 & 54.20 \\
  & DYPSA & 96.12 & 2.24 & 1.64 & 0.59 & 77.82 \\
APLAWD  & ZFR & \textbf{98.89} & 0.59 & 0.52 & 0.55 & 57.87 \\
  & SEDREAMS & 98.67 & 0.82 & \textbf{0.51} & \textbf{0.45} & 85.15 \\
  & YAGA & 98.88 & \textbf{0.52} & 0.60 & 0.49 & \textbf{85.51} \\
\hline
\end{tabular}
\caption{Summary of the performance of the five methods of GCI estimation for the six databases.}
\label{tab:TabClean}
\end{table*}

In terms of \emph{reliability} performance, SEDREAMS and YAGA algorithms generally give the highest identification rates. Among others, it turns out that SEDREAMS correctly identifies more than 98$\%$ of GCIs for any dataset. This is also true for YAGA, except on the RAB database where it reaches 95.70$\%$. Although the performance of ZFR is below these two techniques for JMK, RAB and KED speakers, its results are rather similar on other datasets, obtaining even the best reliability scores on SLT and APLAWD. As for the DYPSA method, its performance remains behind SEDREAMS and YAGA, albeit it reaches IDRs comprised between 95.54$\%$ and 98.26$\%$, except for the RAB speaker where the technique fails, leading to an important amount of false alarms (15.80$\%$). Finally the HE-based approach is outperformed by all other methods most of the time. However it achieves on all databases identification rates, comprised between 91.74$\%$ and 97.04$\%$.

In terms of \emph{accuracy}, it is observed on all the databases, except for the RAB speaker, that YAGA leads the highest rates of frames for which the timing error is lower than 0.25 ms.  The SEDREAMS algorithm gives almost comparable accuracy performance, just below the accuracy of  YAGA. The DYPSA and HE algorithms, are outperformed by YAGA and SEDREAMS on all datasets.  As it was the case for the reliability results, the accuracy of ZFR strongly depends on the considered speaker. It achieves very good results on the BDL and SLT speakers even though the overall accuracy is rather low especially for the KED corpus.

\begin{figure*}[!ht]
  \centering
  \includegraphics[width=1\textwidth]{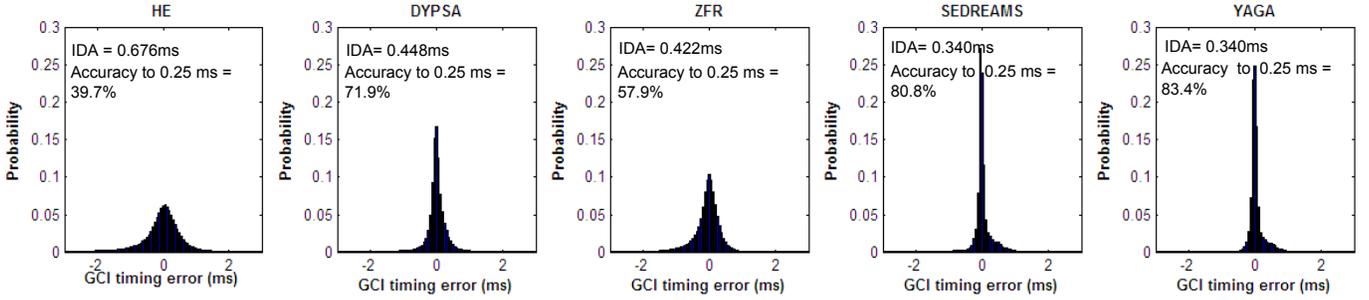}
  \caption{Histograms of the GCI timing error averaged over all databases for the five compared techniques.}
  \label{fig:ErrorHisto}
\end{figure*}

The accuracy performance is illustrated in Fig.~\ref{fig:ErrorHisto} for the five measures. The distributions of the GCI identification error $\xi$ is averaged over all datasets. The histograms for the SEDREAMS and YAGA methods are the sharpest and are highly similar. It is worth pointing out that some discrepancy is expected even if the GCI methods identify the acoustic events with high accuracy, since the delay between the speech signal, recorded by the microphone, and the EGG does not remain constant during recordings.

In conclusion from the results of Table \ref{tab:TabClean}, the SEDREAMS and YAGA techniques, with highly similar performance, generally outperform other methods of GCI detection on clean speech, both in terms of reliability and accuracy. The ZFR method can also reach comparable (or even slightly better) results for some databases, but its performance is observed to be strongly sensitive to the considered speaker. In general, these three approaches are respectively followed by the DYPSA algorithm and the HE-based method.

\subsection{Performance based on Causal-Anticausal Deconvolution}\label{ssec:MixedPhase}

As introduced in Section \ref{sssec:MixedPhase}, the Causal-Anticausal deconvolution is a well-suited approach for evaluating our techniques of GCI determination on a concrete application of speech processing. It was indeed emphasized that this method of glottal flow estimation is highly sensitive to GCI location errors. Besides we presented in Section \ref{sssec:MixedPhase} an objective spectral criterion for deciding whether the mixed-phase separation fails or not. It is important to note at this point that the constraint of precise GCI-synchronization is a necessary, but not sufficient, condition for having a correct deconvolution.

\begin{figure}[!ht]
  \centering
  \includegraphics[width=0.45\textwidth]{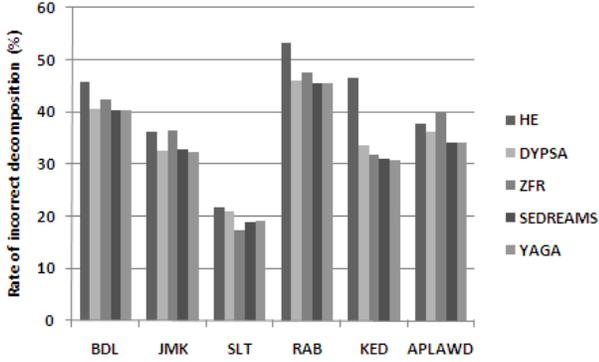}
  \caption{Proportion of speech frames leading to an incorrect mixed-phase deconvolution using all GCI estimation techniques on all databases.}
  \label{fig:Mixed-PhaseResults}
\end{figure}

Figure \ref{fig:Mixed-PhaseResults} displays, for all databases and GCI estimation techniques, the proportion of speech frames that are incorrectly decomposed via mixed-phase separation (achieved in this work by the complex cepstrum-based algorithm \cite{Drugman-CCD}). It can be observed that for all datasets (except for SLT), SEDREAMS and YAGA outperform other approaches and lead again to almost the same results. They are closely followed by the DYPSA algorithm whose accuracy was also shown to be quite high in the previous section. The ZFR method turns out to be generally outperformed by these three latter techniques, but still gives the best results on the SLT voice. Finally, it is seen that the HE-based approach leads to the highest rates of incorrectly decomposed frames. Interestingly, these results achieved in the applicative context of the mixed-phase deconvolution corroborate the conclusions drawn from the comparison with EGG signals, especially regarding their accuracy to $\pm 0.25$ ms (see Section \ref{ssec:compEGG}). This means that the choice of an efficient technique of GCI estimation, as those compared in this work, may significantly improve the performance of applications of speech processing for which a pitch-synchronous analysis or synthesis is required.


\section{Robustness of GCI Extraction Methods}\label{sec:Robustness}
In some speech processing applications, such as speech synthesis, utterances are recorded in well controlled conditions. For such high-quality speech signals, the performance of GCI estimation techniques was studied in Section \ref{sec:ExpClean}. For many other types of speech processing systems however, there is no other choice than capturing the speech signal in a \emph{real world environment}, where noise and/or reverberation may dramatically degrade its quality. The goal of this section is to evaluate how GCI detection methods are affected by additive noise (Section \ref{ssec:Noise}) and by reverberation (Section \ref{ssec:Reverb}). Note that results presented here below were averaged over the six databases.

\subsection{Robustness to an Additive Noise}\label{ssec:Noise}
\begin{figure*}[!ht]
  \centering
  \includegraphics[width=1\textwidth]{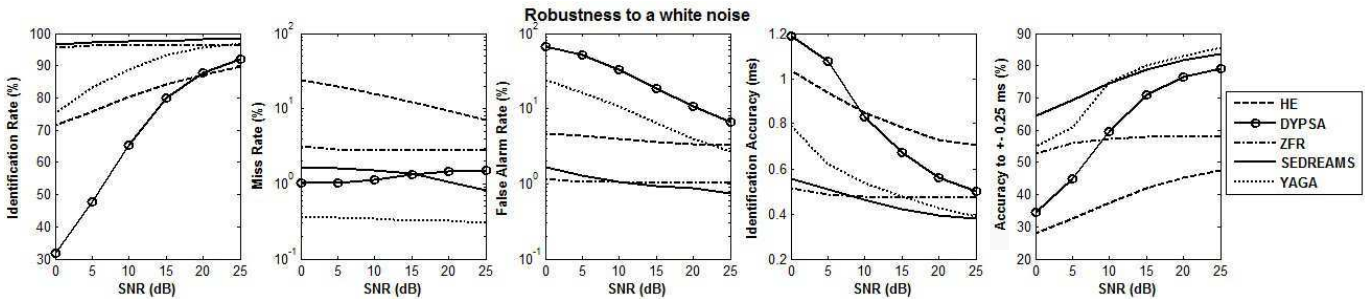}
  \caption{Robustness of GCI estimation methods to an additive white noise, according to the five measures of performance. Miss rate and false alarm rate are in logarithmic scale.}
  \label{fig:Robustness_WhiteNoise}
\end{figure*}

\begin{figure*}[!ht]
  \centering
  \includegraphics[width=1\textwidth]{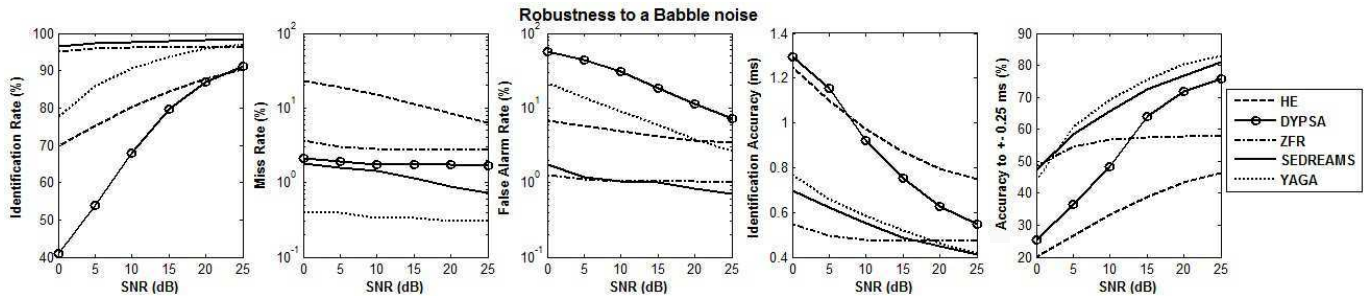}
  \caption{Robustness of GCI estimation methods to an additive babble noise, according to the five measures of performance. Miss rate and false alarm rate are in logarithmic scale.}
  \label{fig:Robustness_BabbleNoise}
\end{figure*}

In a first experiment, noise was added to the original speech waveform at various Signal-to-Noise Ratio (SNR). Both a White Gaussian Noise (WGN) and a babble noise (also known as cocktail party noise) were considered. The noise signals were taken from the Noisex-92 database \cite{Noisex}, and were added so as to control the segmental SNR without silence removal. Results for these two noise types are exhibited in Figs. \ref{fig:Robustness_WhiteNoise} and \ref{fig:Robustness_BabbleNoise} according to the measures detailed in Section \ref{sssec:compEGG}. In these figures, miss rate and false alarm rate are in logarithmic scale for the sake of clarity. It is observed that, for both noise types, the general trends remain unchanged. However it turns out that the degradation of reliability is more severe with the white noise, while the accuracy is more affected by the babble noise.

In terms of reliability, it is noticed that SEDREAMS and ZFR lead to the best robustness, since their performance is almost unchanged up to 0dB of SNR. Secondly, the degradation for YAGA and HE is almost equivalent, while it is noticed that DYPSA is strongly affected by additive noise. Among others, it is observed that HE is characterized by an increasing miss rate as the noise level increases, while the degradation is reflected by an increasing number of false alarms for DYPSA, and for YAGA in a lesser extent. This latter observation is probably due to the difficulty of the dynamic programing process to deal with spurious GCI candidates caused by the additive noise. 

Regarding the accuracy capabilities, similar conclusions hold. Nevertheless the sensitivity of SEDREAMS is this time comparable to that of YAGA and HE. Again, the ZFR algorithm is found to be the most robust technique, while DYPSA is the one presenting the strongest degradation and HE displays the worst identification accuracy.

Good results of robustness for ZFR and SEDREAMS can be explained by the low sensitivity to an additive noise of respectively the zero-frequency resonators and the mean-based signal. In the case of ZFR, analysis is confined around 0 Hz, which tends to minimize not only the effect of the vocal tract, but of an additive noise as well. As for SEDREAMS, the mean-based signal is computed as in Equation \ref{eq:Mean}, which is a linear relation. In other words, the mean-based signal of the noise is added to the mean-based signal of the speech signal. On a duration of 1.75$\cdot$$\bar{T}_{0,mean}$, the white noise is assumed to be almost zero-mean. A similar conclusion is observed for the babble noise, which is composed of several sources of speech talking at the same time. It can indeed be understood that the higher the number of sources in the babble noise, the lesser its degradation on the target mean-based signal. Finally, the strong sensitivity of DYPSA and YAGA might be explained, among others, by the fact that they rely on some thresholds, which have been optimized for clean speech.

\subsection{Robustness to Reverberation}\label{ssec:Reverb}

In many modern telecommunication applications, speech signals are obtained in enclosed spaces with the talker situated at a distance from the microphone. The received speech signal is distorted by reverberation, caused by reflected signals from walls and hard objects, diminishing intelligibility and perceived speech quality~\cite{Bolt1949, Kuttruff2000}. It has been further observed that the performance of GCI identification algorithms is degraded when applied to reverberant signals~\cite{Gaubitch2007}. 

The observation of reverberant speech at microphone $m$ is
\begin{equation}\label{eq:revsp1}
    x_m(n) = h_m(n)\ast {s}(n), \quad m=1,2,\ldots,M,
\end{equation}
where $h_m(n)$ is the $L$-tap Room Impulse Response (RIR) of the acoustic channel between the source to the $m$th microphone. It has been shown that multiple time-aligned observations with a microphone array can be exploited for GCI estimation in reverberant environments~\cite{Thomas2007a}; in this paper we only consider the robustness of single-channel algorithms to the observation at channel $x_1(n)$. RIRs are characterised by the value $T_{60}$, defined as the time for the amplitude of the RIR to decay to -60dB of its initial value. A room measuring 3x4x5~m and $T_{60}$ ranging \{100, 200,~\ldots,~500\}~ms was simulated using the source-image method~\cite{Allen1979} and the simulated impulse responses convolved with the clean speech signals described in Section~\ref{sec:Assessment}.

\begin{figure*}[!ht]
  \centering
  \includegraphics[width=1\textwidth]{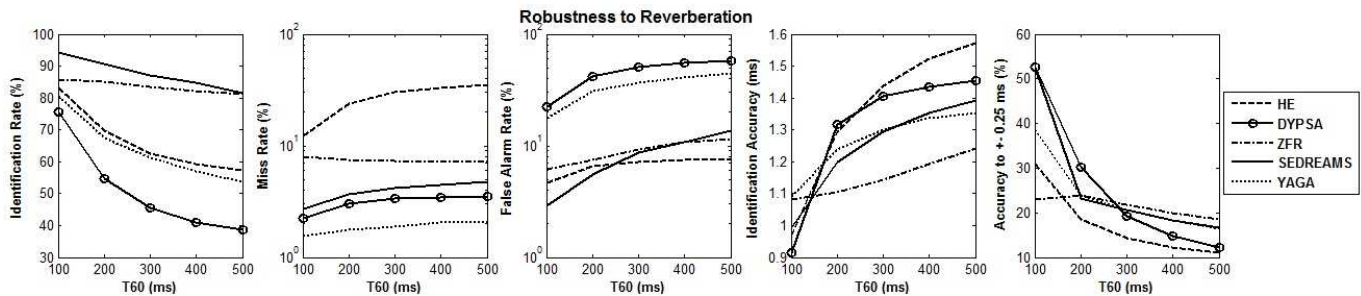}
  \caption{Robustness of GCI estimation methods to reverberation, according to the five measures of performance. Miss rate and false alarm rate are in logarithmic scale.}
  \label{fig:Robustness_Reverb}
\end{figure*}

The results in Figure~\ref{fig:Robustness_Reverb} show that the performance of the algorithms monotonically reduces with increasing reverberation, with the most significant change in performance occurring between $T_{60}=100$ and $200$~ms. They also reveal that reverberation has a particularly detrimental effect upon identification rate of the LP-based approaches, namely HE, DYPSA and YAGA. This is consistent with previous studies which have shown that the RIR results in additional spurious peaks in the LP residual of similar amplitude to the voiced excitation~\cite{Brandstein2001,Yegnanarayana2000}, generally increasing false alarm rate for DYPSA and YAGA but increasing miss rate for HE. Although spurious peaks result in increased false alarms, the identification accuracy of the hits is much less affected. The non-LP approaches generally exhibit better identification rates in reverberation, in particular SEDREAMS. The ZFR algorithm appears to be the least sensitive to reverberation while providing the best overall performance. However, the challenge of GCI detection from single-channel reverberant observations remains an ongoing research problem as no single algorithm consistently provides good results for all five measures.

\section{Computational Complexity of GCI Extraction Methods}\label{sec:Complexity}

In the previous sections, methods of GCI estimation have been compared according to their reliability and accuracy both in clean conditions (Section \ref{sec:ExpClean}) and noisy/reverberant environments (Section \ref{sec:Robustness}). In order to provide a complete comparison, an investigation into computational complexity is described in this section. The algorithms described in Section \ref{sec:Methods} are relatively complex and their computational complexity is highly data-dependent; it is therefore difficult to find a closed-form expression for computational complexity. In this section we discuss those components that present a high computational load and provide a quantitative analysis based upon empirical measurements.


For HE, ZFR and SEDREAMS, the most time-consuming step is the computation of the oscillating signal which they rely on. For the HE method, the CoG-based signal is computed from Equation \ref{eq:CoG} and requires, for each sample, around $2.2\cdot F_s/\bar{T}_{0,mean}$ multiplications and the same number of additions. For ZFR, the mean removal operation (Equation \ref{eq:MeanRemoval}) is repeated three times, and thus requires about $4.5\cdot F_s/\bar{T}_{0,mean}$ additions for each sample of the zero frequency-filtered signal. As for the SEDREAMS algorithm, the computation of each sample of the mean-based signal (Equation \ref{eq:Mean}) requires $1.75\cdot F_s/\bar{T}_{0,mean}$ multiplications and the same number of additions. 

However, it is worth emphasizing that the computation time requested by HE and SEDREAMS can be significantly reduced. Indeed these methods only exploit some particular points of the oscillating signal they rely on: the negative zero-crossings for HE, and the extrema for SEDREAMS. It is then not necessary to compute all the samples of these signals for finding these particular events. Based on this idea, a multiscale approach can be used. For example, the oscillating signals can be first calculated only for the samples multiple of $2^p$. From this downsampled signal, a first approximation of the particular points is obtained. This approximation is then refined iteratively using the $p$ successive smaller scales. The lower bounding value of $p$ means there are, for the first approximation, at least two samples per cycle. In the following, we used $p=4$ so that voices with pitch up to 570 Hz can be processed. The resulting methods are hereafter called \emph{Fast HE} and \emph{Fast SEDREAMS}. Notice that a similar acceleration cannot be transposed to ZFR as the operation of mean removal is applied 3 times successively.

In the case of DYPSA and YAGA, the signal conditioning stages present a relatively low computational load. The LPC residual, Group Delay Function and Multiscale Product scale approximately $\mathcal{O}(N^2)$, $\mathcal{O}(N\log_2 N)$ and $\mathcal{O}(N)$ respectively, where $N$ is the total number of samples in the speech signal. Computational load is significantly heavier in the dynamic programming stages due to the large number of erroneous GCI candidates that must be removed. In particular, the waveform similarity measure, used to determine the similarity of two neighbouring cycles, presents a high computational load due to the large number of executions required to find the optimum path. At present this is calculated on full-band speech although it is expected that calculation of waveform similarity on a downsampled signal may yield similar results for a much-reduced computational load. A second optimization lies in the length of the group delay evaluation window, which is inversely proportional to the number of candidates generated. At present this takes a fixed value based upon the maximum expected $f_0$; far fewer erroneous candidates could be generated by dynamically varying the length based upon a crude initial estimate of $f_0$.

So as to compare their computational complexity, the \emph{Relative Computation Time} (RCT) of each GCI estimation method is evaluated on all databases:

\begin{equation}\label{eq:RCT}
RCT (\%) = 100\cdot \frac{\text{CPU time (s)}}{\text{Sound duration (s)}}
\end{equation}

Table \ref{tab:TabComplexity} shows, for both male and female speakers, the averaged RCT obtained for our Matlab implementations and with a Intel Core 2 Duo T7500 2.20 GHz CPU with 3GB of RAM. First of all, it is observed that results are ostensibly the same for both genders. Regarding the non-accelerated versions of the GCI detection methods, it turns out that DYPSA is the fastest (with a RCT around 20\%), followed by SEDREAMS and YAGA, which both have a RCT of about 28\%. The HE-based technique gives a RCT of around 33\%, and ZFR, due to its operation of mean removal which has to be repeated three times, is the slowest method with a RCT of 75\%. Interestingly, it is noticed that the accelerated versions of HE and SEDREAMS reduce the computation time by about 5 times on male voices, and by around 4 times for female speakers. This leads to the fastest GCI detection algorithms, reaching a RCT of around 6\% for Fast SEDREAMS, and about 8\% for Fast HE. Note finally that these results could be highly reduced by using, for example, a C-implementation of these techniques, albeit the conclusions remain identical.

\begin{table}[!ht]
\centering
\begin{tabular}{| c || c | c |}
\hline
\textbf{Method} & \textbf{Male} & \textbf{Female}\\
\hline
\hline
HE & 35.0 & 31.8\\
\hline
Fast HE & 7.6 & 7.8\\
\hline
DYPSA & 19.9 & 19.4\\
\hline
ZFR & 75.7 & 74.9\\
\hline
SEDREAMS & 27.8 & 27.1\\
\hline
Fast SEDREAMS & 5.4 & 6.9\\
\hline
YAGA & 28.6 & 28.3\\
\hline
\end{tabular}
\caption{Relative Computation Time (RCT), in \%, for all methods and for male and female speakers. Results have been averaged across all databases.}
\label{tab:TabComplexity}
\end{table}

\section{Conclusion}\label{sec:conclu}
This paper gave a comparative evaluation of five of the most effective methods for automatically determining GCIs from the speech waveform: Hilbert Envelope-based detection (HE), the Zero Frequency Resonator-based method (ZFR), DYPSA, SEDREAMS and YAGA. The performance of these methods was assessed on six databases containing several male and female speakers, for a total amount of data of approximately four hours. In our first experiments on clean speech, the SEDREAMS and YAGA algorithms gave the best results, with a comparable performance. For \emph{any} database, they reached an identification rate greater than $98\%$ and more than $80\%$ of GCIs were located with an accuracy of $0.25$ ms. Although the ZFR technique can lead to a similar performance, its efficiency can also be rather low in some cases. In general, these three approaches were shown to respectively outperform DYPSA and HE. In a second experiment on clean speech, the impact of the performance of these five methods was studied on a concrete application of speech processing: the causal-anticausal deconvolution. Results showed that adopting a GCI detection with high performance could significantly improve the proportion of correctly deconvolved frames. In the last experiment, the robustness of the five techniques to additive noise, as well as to reverberation was investigated. The ZFR and SEDREAMS algorithms were shown to have the highest robustness, with an almost unchanged reliability. DYPSA was observed to be especially affected, which was reflected by a high rate of false alarms. Although the degradation of accuracy was relatively slow with the level of additive noise, it was noticed that reverberation dramatically affects the precision GCI detection methods. In addition, the computational complexity of the algorithms was studied. A method for accelerating the GCI location using HE and SEDREAMS was proposed. This led, for our Matlab implementation, to a computation time about 6\% real-time for the fast version of SEDREAMS. 

Depending on the speech application to design, some GCI methods could be preferred to some others, based on their performance for the criteria studied in this article. However, if the application is placed in an unknown environment, we suggest the use of SEDREAMS for the following reasons: \emph{i)} it gave the best results with YAGA on clean speech, \emph{ii)} it was the best performing technique in noisy conditions, \emph{iii)} it led with ZFR to the best robustness in a reverberant environment, and \emph{iv)} it was the most suited method for a real-time implementation.


%



\section*{Acknowledgment}

Thomas Drugman is supported by the Belgian Fonds National de la Recherche Scientifique (FNRS). Authors also would like to thank the reviewers for their fruitful comments.

\ifCLASSOPTIONcaptionsoff
  \newpage
\fi



%
\bibliographystyle{IEEEtran}
\bibliography{bare_jrnl}

\begin{thebibliography}{10}
\providecommand{\url}[1]{#1}
\csname url@samestyle\endcsname
\providecommand{\newblock}{\relax}
\providecommand{\bibinfo}[2]{#2}
\providecommand{\BIBentrySTDinterwordspacing}{\spaceskip=0pt\relax}
\providecommand{\BIBentryALTinterwordstretchfactor}{4}
\providecommand{\BIBentryALTinterwordspacing}{\spaceskip=\fontdimen2\font plus
\BIBentryALTinterwordstretchfactor\fontdimen3\font minus
  \fontdimen4\font\relax}
\providecommand{\BIBforeignlanguage}[2]{{%
\expandafter\ifx\csname l@#1\endcsname\relax
\typeout{** WARNING: IEEEtran.bst: No hyphenation pattern has been}%
\typeout{** loaded for the language `#1'. Using the pattern for}%
\typeout{** the default language instead.}%
\else
\language=\csname l@#1\endcsname
\fi
#2}}
\providecommand{\BIBdecl}{\relax}
\BIBdecl

\bibitem{Catford1977}
J.~C. Catford, \emph{Fundamental Problems in Phonetics}.\hskip 1em plus 0.5em
  minus 0.4em\relax Indiana University Press, 1977.

\bibitem{ITU_T_P862}
\emph{Perceptual evaluation of speech quality ({PESQ}), an objective method for
  end-to-end speech quality assessment of narrowband telephone networks and
  speech codecs}, International Telecommunications Union ({ITU-T})
  Recommendation P.862, Feb. 2001.

\bibitem{Moulines1990}
E.~Moulines and F.~Charpentier, ``Pitch-synchronous waveform processing
  techniques for text-to-speech synthesis using diphones,'' \emph{Speech
  Communication}, vol.~9, no. 5--6, pp. 453--467, Dec. 1990.

\bibitem{Gaubitch2007}
N.~D. Gaubitch and P.~A. Naylor, ``Spatiotemporal averaging method for
  enhancement of reverberant speech,'' in \emph{Proc. {IEEE} Intl. Conf.
  Digital Signal Processing (DSP)}, Cardiff, UK, 2007.

\bibitem{Wong1979}
D.~Y. Wong, J.~D. Markel, and J.~A.~H. Gray, ``Least squares glottal inverse
  filtering from the acoustic speech waveform,'' \emph{{IEEE} Trans. Acoust.,
  Speech, Signal Process.}, vol.~27, no.~4, pp. 350--355, Aug. 1979.

\bibitem{HNM}
Y.~Stylianou, ``Applying the harmonic plus noise model in concatenative speech
  synthesis,'' \emph{{IEEE} Trans. Speech Audio Process.}, vol.~9, pp. 21--29,
  2001.

\bibitem{DSM}
T.~Drugman, G.~Wilfart, and T.~Dutoit, ``A deterministic plus stochastic model
  of the residual signal for improved parametric speech synthesis,'' in
  \emph{Proc. Interspeech Conference}, 2009.

\bibitem{Thomas2009}
M.~R.~P. Thomas, J.~Gudnason, and P.~A. Naylor, ``Data-driven voice source
  waveform modelling,'' in \emph{Proc. {IEEE} Intl. Conf. on Acoustics, Speech
  and Signal Processing (ICASSP)}, Taipei, Taiwan, Apr. 2009.

\bibitem{MixedPhase}
B.~Bozkurt and T.~Dutoit, ``Mixed-phase speech modeling and formant estimation,
  using differential phase spectrums,'' in \emph{ISCA ITRW VOQUAL03}, 2003, pp.
  21--24.

\bibitem{Strube1974a}
H.~W. Strube, ``Determination of the instant of glottal closure from the speech
  wave,'' \emph{J. Acoust. Soc. Am.}, vol.~56, no.~5, pp. 1625--1629, 1974.

\bibitem{Rabiner1988}
L.~Rabiner and R.~Schafer, \emph{Digital Processing of Speech Signals}.\hskip
  1em plus 0.5em minus 0.4em\relax New Jersey: Prentice Hall, 1988.

\bibitem{Makhoul1975}
J.~Makhoul, ``Linear prediction: A tutorial review,'' \emph{Proc. {IEEE}},
  vol.~63, no.~4, pp. 561--580, Apr. 1975.

\bibitem{Plumpe1999}
M.~D. Plumpe, T.~F. Quatieri, and D.~A. Reynolds, ``Modeling of the glottal
  flow derivative waveform with application to speaker identification,''
  \emph{{IEEE} Trans. Speech Audio Process.}, vol.~7, no.~5, pp. 569--576, Sep.
  1999.

\bibitem{Naylor2007a}
P.~A. Naylor, A.~Kounoudes, J.~Gudnason, and M.~Brookes, ``Estimation of
  glottal closure instants in voiced speech using the {DYPSA} algorithm,''
  \emph{{IEEE} Trans. Speech Audio Process.}, vol.~15, no.~1, pp. 34--43, 2007.

\bibitem{Thomas2010b}
M.~R.~P. Thomas, J.~Gudnason, and P.~A. Naylor, ``Detection of glottal opening
  and closing instants in voiced speech using the {YAGA} algorithm,''
  \emph{Submitted for peer review}, 2010.

\bibitem{Chytil2006}
P.~Chytil and M.~Pavel, ``Variability of glottal pulse estimation using
  cepstral method,'' in \emph{Proc. 7th Nordic Signal Processing Symposium
  ({NORSIG})}, 2006, pp. 314--317.

\bibitem{Thomas2007a}
M.~R.~P. Thomas, N.~D. Gaubitch, and P.~A. Naylor, ``Multichannel {DYPSA} for
  estimation of glottal closure instants in reverberant speech,'' in
  \emph{Proc. European Signal Processing Conf. (EUSIPCO)}, 2007.

\bibitem{Ananthapadmanabha1979}
T.~V. Ananthapadmanabha and B.~Yegnanarayana, ``Epoch extraction from linear
  prediction residual for identification of closed glottis interval,''
  \emph{{IEEE} Trans. Acoust., Speech, Signal Process.}, vol.~27, pp. 309--319,
  1979.

\bibitem{Ma1994}
C.~Ma, Y.~Kamp, and L.~F. Willems, ``A {Frobenius} norm approach to glottal
  closure detection from the speech signal,'' \emph{{IEEE} Trans. Speech Audio
  Process.}, vol.~2, pp. 258--265, Apr. 1994.

\bibitem{Murty2008}
K.~S.~R. Murty and B.~Yegnanarayana, ``Epoch extraction from speech signals,''
  \emph{{IEEE} Trans. Audio, Speech, Lang. Process.}, vol.~16, no.~8, pp.
  1602--1613, 2008.

\bibitem{SEDREAMS}
T.~Drugman and T.~Dutoit, ``Glottal closure and opening instant detection from
  speech signals,'' in \emph{Proc. Interspeech Conference}, 2009.

\bibitem{Bouzid2004}
A.~Bouzid and N.~Ellouze, ``Glottal opening instant detection from speech
  signal,'' in \emph{Proc. European Signal Processing Conf. (EUSIPCO)}, Vienna,
  Sep. 2004, pp. 729--732.

\bibitem{Tuan1999a}
V.~N. Tuan and C.~{d'Alessandro}, ``Robust glottal closure detection using the
  wavelet transform,'' in \emph{Eurospeech}, Budapest, Sep. 1999, pp.
  2805--2808.

\bibitem{Polarity}
W.~Ding and N.~Campbell, ``Determining polarity of speech signals based on
  gradient of spurious glottal waveforms,'' \emph{Proc. {IEEE} Intl. Conf. on
  Acoustics, Speech and Signal Processing (ICASSP)}, vol.~2, pp. 857--860,
  1998.

\bibitem{Yegnanarayana1979}
T.~Ananthapadmanabha and B.~Yegnanarayana, ``Epoch extraction from linear
  prediction residual for identification of closed glottis interval,''
  \emph{IEEE Trans. Acoust. Speech Signal Proc.}, vol.~27, pp. 309--319, 1979.

\bibitem{Cheng1989}
Y.~M. Cheng and D.~O'Shaughnessy, ``Automatic and reliable estimation of
  glottal closure instant and period,'' \emph{{IEEE} Trans. Acoust., Speech,
  Signal Process.}, vol.~37, pp. 1805--1815, Dec. 1989.

\bibitem{Rao2007}
K.~S. Rao, S.~R.~M. Prasanna, and B.~Yegnanarayana, ``Determination of instants
  of significant excitation in speech using {H}ilbert envelope and group delay
  function,'' \emph{{IEEE} Signal Process. Lett.}, vol.~14, no.~10, pp.
  762--765, 2007.

\bibitem{Yegnanarayana1995}
B.~Yegnanarayana and R.~Smits, ``A robust method for determining instants of
  major excitations in voiced speech,'' in \emph{Proc. {IEEE} Intl. Conf. on
  Acoustics, Speech and Signal Processing (ICASSP)}, 1995, pp. 776--779.

\bibitem{Brookes2006}
M.~Brookes, P.~A. Naylor, and J.~Gudnason, ``A quantitative assessment of group
  delay methods for identifying glottal closures in voiced speech,''
  \emph{{IEEE} Trans. Speech Audio Process.}, vol.~14, 2006.

\bibitem{Mallat1992}
S.~Mallat and S.~Zhong, ``Characterization of signals from multiscale edges,''
  \emph{{IEEE} Trans. Pattern Anal. Mach. Intell.}, vol.~14, no.~7, pp.
  710--732, 1992.

\bibitem{Sadler1999}
B.~M. Sadler and A.~Swami, ``Analysis of multiscale products for step detection
  and estimation,'' \emph{{IEEE} Trans. Inf. Theory}, vol.~45, no.~3, pp.
  1043--1051, 1999.

\bibitem{Festvox}
[Online], ``The festvox website,'' in \emph{\emph{http://festvox.org/}}.

\bibitem{Lindsey1987}
G.~Lindsey, A.~Breen, and S.~Nevard, ``{SPAR}'s archivable actual-word
  databases,'' University College London, Technical Report, 1987.

\bibitem{Hunt1978}
M.~J. Hunt, ``Automatic correction of low-frequency phase distortion in
  analogue magnetic recordings,'' \emph{Acoustics Letters}, vol.~2, pp. 6--10,
  1978.

\bibitem{Funaki1999}
K.~Funaki and K.~T. Y.~Miyanaga, ``Recursive {ARMAX} speech analysis based on a
  glottal source model with phase compensation,'' \emph{Signal Processing},
  vol.~74, no.~3, pp. 279--295, May 1999.

\bibitem{Doval-CALM}
B.~Doval, C.~d'Alessandro, and N.~Henrich, ``The voice source as a
  causal/anticausal linear filter,'' in \emph{ISCA ITRW VOQUAL03}, 2003, pp.
  15--19.

\bibitem{ZZT}
B.~Bozkurt, B.~Doval, C.~d'Alessandro, and T.~Dutoit, ``Zeros of z-transform
  representation with application to source-filter separation in speech,''
  \emph{IEEE Signal Processing Letters}, vol.~12, 2005.

\bibitem{Drugman-CCD}
T.~Drugman, B.~Bozkurt, and T.~Dutoit, ``Complex cepstrum-based decomposition
  of speech for glottal source estimation,'' in \emph{Proc. Interspeech
  Conference}, 2009.

\bibitem{Fant1985}
G.~Fant, J.~Liljencrants, and Q.~Lin, ``A four-parameter model of glottal
  flow,'' \emph{STL-QPSR}, vol.~26, no.~4, pp. 1--13, 1985.

\bibitem{Noisex}
[Online], ``Noisex-92,'' in \emph{http://www.speech.cs.
  cmu.edu/comp.speech/Sectionl/Data/noisex.html}.

\bibitem{Bolt1949}
R.~H. Bolt and A.~D. MacDonald, ``Theory of speech masking by reverberation,''
  \emph{J. Acoust. Soc. Am.}, vol.~21, pp. 577--580, 1949.

\bibitem{Kuttruff2000}
H.~Kuttruff, \emph{Room Acoustics}, 4th~ed.\hskip 1em plus 0.5em minus
  0.4em\relax Taylor \& Frances, 2000.

\bibitem{Allen1979}
J.~B. Allen and D.~A. Berkley, ``Image method for efficiently simulating
  small-room acoustics,'' \emph{J. Acoust. Soc. Am.}, vol.~65, no.~4, pp.
  943--950, 1979.

\bibitem{Brandstein2001}
M.~S. Brandstein and D.~B. Ward, Eds., \emph{Microphone Arrays: Signal
  Processing Techniques and Applications}.\hskip 1em plus 0.5em minus
  0.4em\relax Springer-Verlag, 2001.

\bibitem{Yegnanarayana2000}
B.~Yegnanarayana and P.~Satyanarayana, ``Enhancement of reverberant speech
  using {LP} residual signal,'' \emph{{IEEE} Trans. Speech Audio Process.},
  vol.~8, no.~3, pp. 267--281, 2000.

\end{thebibliography}

\end{document}